\documentclass[journal]{IEEEtran}
\IEEEoverridecommandlockouts
\usepackage{cite}
\usepackage{amsmath,amssymb,amsfonts}
\usepackage{bm}
\usepackage{algorithm}
\usepackage{algorithmic}
\usepackage{graphicx}
\usepackage{textcomp}  
\usepackage{xcolor}
\usepackage{subfigure}
\usepackage{authblk}
\usepackage{lettrine}

\def\BibTeX{{\rm B\kern-.05em{\sc i\kern-.025em b}\kern-.08em
    T\kern-.1667em\lower.7ex\hbox{E}\kern-.125emX}}
\begin{document}

\title{Horizontal and Vertical Collaboration for VR Delivery in MEC-Enabled Small-Cell Networks  }

\author{Zhuojia~Gu, 
        Hancheng~Lu,~\IEEEmembership{Senior Member,~IEEE,}
        and Chenkai Zou

}

\maketitle

\begin{abstract}
Due to the large bandwidth, low latency and computationally intensive features of virtual reality (VR) video applications, the current resource-constrained wireless and edge networks cannot meet the requirements of on-demand VR delivery. In this letter, we propose a joint horizontal and vertical collaboration architecture in mobile edge computing (MEC)-enabled small-cell networks for VR delivery. In the proposed architecture, multiple MEC servers can jointly provide VR head-mounted devices (HMDs) with edge caching and viewpoint computation services, while the computation tasks can also be performed at HMDs or on the cloud. Power allocation at base stations (BSs) is considered in coordination with horizontal collaboration (HC) and vertical collaboration (VC) of MEC servers to obtain lower end-to-end latency of VR delivery.
A joint caching, power allocation and task offloading problem is then formulated, and a discrete branch-reduce-and-bound (DBRB) algorithm inspired by monotone optimization is proposed to effectively solve the problem.
Simulation results demonstrate the advantage of the proposed architecture and algorithm in terms of existing ones.
\end{abstract}

\begin{IEEEkeywords}
VR delivery, horizontal and vertical collaboration, MEC, end-to-end latency.
\end{IEEEkeywords}

\section{introduction}
\IEEEPARstart{O}{n-demand} virtual reality (VR) applications require ultra-large bandwidth, ultra-low latency and abundant computation resources in delivery networks, which hinders its widespread popularity in existing 5G networks \cite{hosseini2017view}. By providing caching and computation capabilities at the network edge, mobile edge computing (MEC) can significantly reduce the transmission and computation delays, which are the main components of the end-to-end latency of VR delivery \cite{liu2021learning, qian2020rein}.

Existing works on MEC can be categorized into two types: 1) vertical collaboration (VC)-based MEC, and 2) horizontal collaboration (HC)-based MEC. The VC-based MEC typically adopts a three-tier hierarchical architecture including the terminal tier, the edge tier, and the cloud tier to conduct collaborative caching and computation offloading strategies \cite{abbas2018mobile}. In the scenario of VR delivery, the authors in \cite{sun2019communications, dang2019joint} proposed the joint communications, caching and computation (3C) modeling of a hierarchical VR delivery network architecture, and revealed the 3C resources trade-off for saving communication bandwidth while meeting the low latency requirement.
On the other hand, considering that mobile edge servers (MESs) are usually deployed sinking to the densely deployed base stations (BSs) in 5G wireless networks \cite{wu2019delay, gu2021association}, HC-based MEC has been proposed for higher utilization efficiency of edge resources, facilitated by the higher speed wireless connections in dense small-cell networks than that of the backhaul link \cite{wang2018enabling}. The authors in \cite{saputra2021anovel, dai2020a} investigated the HC strategies of caching or computation resources among MESs, respectively, and the performance gains using HC in terms of the service delay were validated.



From the discussion above, it is evident that both VC and HC are effective methods to improve the performance of delay-sensitive applications in wireless networks. Intuitively, combining HC with VC in MEC-enabled small-cell networks is a potential solution to further meet the stringent latency requirement of computation-intensive VR applications.
However, there is a lack of studies focusing on the joint HC and VC-assisted MEC for VR video delivery. Considering the specific characteristic of VR video delivery, the design of joint HC and VC in MEC-enabled small-cell networks is non-trivial.
First, since the VR field of views (FoVs) needs to be projected and rendered from monocular videos (MVs) into stereoscopic videos (SVs) before it can be presented at VR head-mounted displays (HMDs), caching MVs in MESs can save the cache capacity, but at the cost of consuming more computation resources. Considering limited caching and computation resources in MESs, it is a critical issue to design HC-based caching and task offloading strategies among MESs to reduce the transmission and computation delays while considering the VC task offloading aspects.
Second, since HC and VC are to be considered in collaboration to further improve the end-to-end latency performance for VR delivery, it will incur a trade-off issue between HC and VC for collaborative caching and task offloading.
In addition, power allocation in dense small-cell networks has a great impact on the transmission rates of wireless links. In the case of limited power resources at BSs, power allocation should be reconsidered in coordination with joint HC and VC-based caching and task offloading strategies to achieve lower total latency for VR delivery.

In this letter, we propose a joint HC and VC architecture in MEC-enabled small-cell networks for VR delivery. A novel task offloading problem with consideration of collaborative caching and power allocation is formulated with the aim of minimizing the end-to-end latency for VR delivery. A joint caching, power allocation and task offloading algorithm envisioned by discrete monotone optimization is proposed to effectively solve the problem. Numerical simulations have been conducted to justify the effectiveness of the joint HC and VC architecture for VR delivery and show that the proposed JCPT algorithm outperforms the benchmark algorithms.

\section{System Model} \label{System_Model}
\subsection{Network Architecture}

\begin{figure}[htbp]
	\centering
	\includegraphics[width=0.48\textwidth]{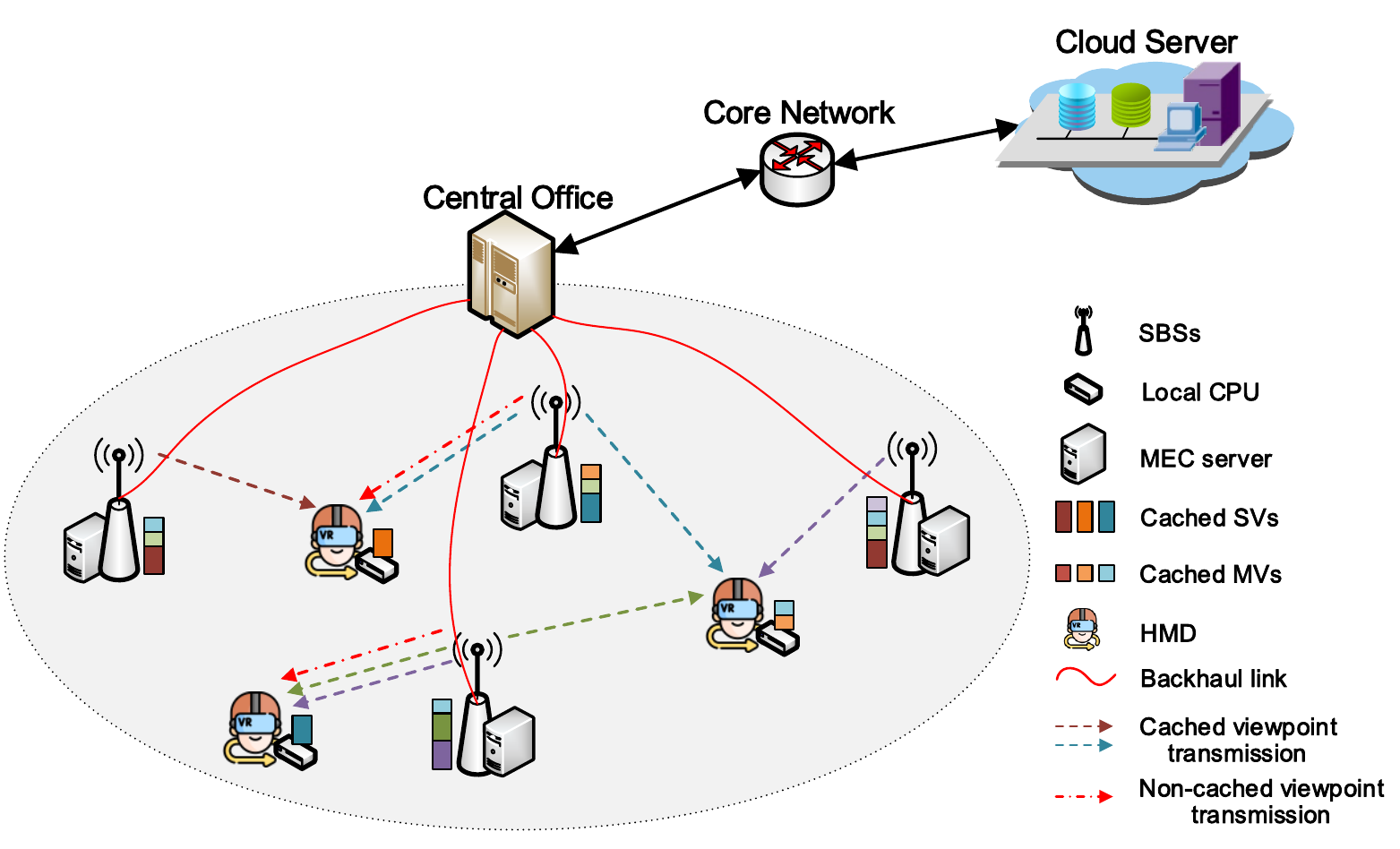}
	\caption{A joint horizontal and vertical collaboration architecture in MEC-enabled small-cell networks for VR delivery.}
	\label{scenario}
\end{figure}

The proposed network architecture is shown in Fig. \ref {scenario}. There are $M$ small-cell base stations (SBSs) equipped with MEC servers (MESs) in the network denoted by an index set $\mathcal{M} = \{1, ..., M\}$, with each MES $m \in \mathcal{M}$.
The VR head-mounted devices (HMDs) are denoted by an index set $\mathcal{U} = \{1, ..., U\}$, with each HMD $u \in \mathcal{U}$.
The set of all viewpoints of VR videos is denoted by $\mathcal{N} = \{1, ..., N\}$, and each viewpoint $i \in \mathcal{N}$ corresponds to a MV and a SV. MVs should be projected and rendered into SVs by using the computation resources at MESs or HMDs, thus the size of the SV version of viewpoint $i$ should be $\alpha (\alpha > 2)$ times larger than that of the corresponding MV version.

Due to the densification of small-cell networks, a HMD can be within the coverage of multiple SBSs, thus the viewpoints can be retrieved from multiple SBSs caching the required MVs or SVs, and the computation task of projecting MVs into SVs can also be processed on multiple MESs. The multiple connections of a HMD to SBSs can be realized by adopting the multi-connectivity technology proposed in 3GPP specification \cite{3GPP, wolf2019how}.
Let $T_{imu}^{M} \in \{0, 1\}$ denote whether the computation task of the $i$-th MV on HMD $u$ is offloaded to MES $m$. Similarly, let $T_{imu}^{C}$ denote whether the computation task of the $i$-th MV on HMD $u$ is offloaded to the cloud service via MES $m$.
Assume that the offloading task for each viewpoint is performed at one server, thus we have
\begin{equation}\label{eq1}
     T_{imu}^{M} + T_{imu}^{C} \le 1, \forall i \in \mathcal{N}, \forall m \in \mathcal{M}, \forall u \in \mathcal{U},
\end{equation}
\begin{equation}
    \sum_{m=1}^{M} T_{imu}^{M} \le 1, \forall i \in \mathcal{N}, \forall u \in \mathcal{U},
\end{equation}
\begin{equation}
    \sum_{m=1}^{M} T_{imu}^{C} \le 1,  \forall i \in \mathcal{N}, \forall u \in \mathcal{U}.
\end{equation}

\subsection{3C Resource Model}

SBSs and HMDs are equipped with caching capacity of $C_u^M$ and $C_u^V$, respectively. Let $c_{im}^{M, M}, c_{im}^{M, S} \in \{0, 1\}$ denote whether the MV or SV version of viewpoint $i$ is cached on MES $m$, respectively. Similarly, let $c_{iu}^{V, M}, c_{iu}^{V, S} \in \{0, 1\}$ denote whether the MV or SV version of viewpoint $i$ is cached on VR HMD $u$. The size of the MV version of viewpoint $i$ is denoted as $d_i$, then the size of the corresponding SV is $\alpha d_i$. Considering the caching capacity constraints at MESs and HMDs, we have
\begin{equation}
    \sum_{i=1}^{N} (c_{im}^{M,M} + \alpha c_{im}^{M, S}) d_i \le C_m^M,  \forall m \in \mathcal{M},
\end{equation}
\vspace{-1mm}
\begin{equation}
	\sum_{i=1}^{N} (c_{iu}^{V, M} + \alpha c_{iu}^{V, S})  d_i \le C_u^V,  \forall u \in \mathcal{U}.
\end{equation}

MESs and VR HMDs are all equipped with computation resources. According to \cite{mao2017a}, if the computation process is executed on an MES, the consumed computation energy to project MV $i$ into SV $i$ can be calculated as
$e_i^{M} = k_M f_M^2 d_i w_i$,
where $k_M$ is the energy efficiency coefficient related to the computing process unit (CPU) architecture of the MES, $f_m$ is the CPU frequency of the MES, and $w_i$ is the required CPU cycles of projection  MV $i$ into  SV $i$ for 1 bit. Then the computation delay on the MES to process MV $i$ can be calculated as $\tau_{i}^M = \frac{d_i w_i}{f_M}$. On the other hand, if the computation process is executed on a VR HMD, the consumed computation energy  is $e_i^{V} = k_V f_V^2 d_i w_i$, and the corresponding computation delay is $\tau_i^V =  \frac{d_i w_i}{f_V}$.
Assume that the requested probabilities of all the viewpoints conforms to the Zipf distribution, i.e., the requested probability of viewpoint $i$ is $q_i = \frac{1 / \mathrm{i}^{\lambda}}{\sum_{\mathrm{j}=1}^{N} 1 / \mathrm{j}^{\lambda}}$, where $\lambda$ reflects the preference skewness of the Zipf distribution. To ensure that the average energy consumption of each MES or HMD does not exceed the threshold $E_{m}^{M}$ or $E_{u}^{V}$, respectively, we have
\begin{equation} \label{equ:5}
    \sum_{u=1}^{U}\sum_{i=1}^{N} q_i (1 -  c_{im}^{M, S}) e_{i}^{M} \mathbb{I}_{\big\{\sum_{m=1}^{M}T_{imu}^{M} = 1\big\}} \le E_{m}^{M} ,  \forall m \in \mathcal{M},
\end{equation}
\begin{equation} \label{equ:4}
     \sum_{i=1}^{N} q_i (1 - c_{iu}^{V, S}) e_{i}^{V} \mathbb{I}_{\big\{ \sum_{m=1}^{M}T_{imu}^{M} + T_{imu}^{C} = 0 \big\}}  \le E_{u}^{V},  \forall u \in \mathcal{U},
\end{equation}
where $\mathbb{I}_{\{\cdot \}}$ is an indicator function, i.e., $\mathbb{I}_{\{z \}} = 1$ if condition $z$ is true, and zero otherwise.

Each SBS allocates transmission power to the associated HMDs, denoted by $p_{mu}$. Assuming universal frequency reuse among SBSs, according to Shannon's formula, the downlink wireless link rate $R_{mu}$ from SBSs $m$ to HMD $u$ can be expressed as
\begin{equation}
    R_{mu}=W \log_2(1+ \gamma_{mu}(\mathbf{p}) ), \label{equ7}
\end{equation}
where $W$ is the bandwidth allocated to HMD $u$, $\gamma_{mu}(\mathbf{p})  \triangleq \frac{p_{mu}  h_{mu}}{I_{Inter-cell}+\ \zeta I_{Intra-cell} + n_{0\ }}$,  $h_{mu}$ is the channel gain between SBS $m$ and HMD $u$, $n_{0}$ is the additive white Gaussian noise (AWGN) power. The inter-cell and intra-cell interferences can be calculated as $I_{Inter-cell} = \sum_{b \neq m}^{b \in M} \sum_{v \neq u}^{v \in U} p_{bv} h_{bv}$ and $I_{Intra-cell} = \sum_{v \neq u}^{v \in U} p_{mv}h_{mv}$, respectively, where $\zeta \in [0,1]$ is the orthogonality factor that represents the capability of intra-cell interference cancelation at the HMD.
Considering the total power constraint $P_T$ at SBS $m$, we have
\begin{equation} \label{eq9}
 \sum_{u=1}^{U} p_{mu}    \le P_T, \forall m \in \mathcal{M}.
\end{equation}

\subsection{HC and VC Task Offloading Model}
A HMD can be associated to multiple SBSs simultaneously to fetch the cached viewpoints and utilize the computation resources on multiple MESs, which is referred to as horizontal collaboration (HC) among MESs. Assume that the end-to-end latency of VR delivery mainly includes the transmission delay and the computation delay. Given the task offloading variable $T_{imu}^{M}$ and the caching status of viewpoint $i$ at MES $m$, the end-to-end latency for HMD $u$ requesting viewpoint $i$ from MES $m$ can be calculated as
\begin{equation}
    \tau_{iu}^M =  \sum_{m = 1}^{M} \bigg(\frac{c_{im}^{M,M} d_i w_i}{f_M} + \frac{\alpha d_i}{R_{mu}} \bigg) T_{imu}^{M}.
\end{equation}
Note that the task can be offloaded to MES $m$ can only when the MV or SV is cached at MES $m$. Thus we have the coupled caching and task offloading constraint,
\begin{equation}\label{coupling-constraint}
    T_{imu}^{M} (1 - c_{im}^{M,M} ) (1 - c_{im}^{M, S} ) < 1, \forall i \in \mathcal{N}, \forall m \in \mathcal{M}, \forall u \in \mathcal{U}.
\end{equation}

On the other hand, computation task can also be implemented locally at the HMD, or be offloaded to cloud server, which is referred to as vertical collaboration (VC). Accordingly, if the computation task is implemented at the HMD, the end-to-end latency just includes the computation delay, which can be calculated as
\begin{equation}
    \tau_{iu}^V =  \mathbb{I}_{\big\{\sum_{m=1}^{M}T_{imu}^{M} + T_{imu}^{C} = 0 \big\}}  \frac{c_{iu}^{V, M}  d_i w_i}{f_V}.
\end{equation}
Note that $\sum_{m=1}^{M}T_{imu}^{M} + T_{imu}^{C} = 0$ can hold only when $c_{iu}^{V, M} = 1$ or $c_{iu}^{V, S} = 1$, i.e., we have
\begin{equation}\label{local-coupling-constraint}
    T_{imu}^{M} + T_{imu}^{C} + c_{iu}^{V, M} + c_{iu}^{V, S} > 0, \forall i \in \mathcal{N}, \forall m \in \mathcal{M}, \forall u \in \mathcal{U}.
\end{equation}
Otherwise, the requested viewpoint should be fetched via the cloud server, and it would incur an additional backhaul retrieving delay $\tau_b$, thus the end-to-end latency can be calculated as
\begin{equation}
	\tau_{iu}^C = \sum_{m = 1}^{M} \bigg( \frac{\alpha\ d_i}{R_{mu}} + \tau_b \bigg) T_{imu}^{C}.
\end{equation}

Taking consideration of collaborated HC and VC of the MESs and the HMDs in the network, the end-to-end latency for HMD $u$ requesting viewpoint $i$ can be written as
\begin{equation}
	\tau_{iu} = \tau_{iu}^M + \tau_{iu}^V + \tau_{iu}^C.
\end{equation}

%

\section{Problem Formulation and Solution}
In this letter, we aim to minimize the average end-to-end latency of VR delivery based on the collaborated HC and VC architecture by jointly considering the caching, power allocation and task offloading strategies. Thus, the proposed optimization problem can be formulated as follows,
\begin{subequations}\label{P1}
\begin{align}
	 \min_{\substack{  \{c_{iu}^{V, M}, c_{iu}^{V, S}, c_{im}^{M, M}, c_{im}^{M, S}, \\  p_{mu}, T_{imu}^{M}, T_{imu}^{C} \}} } \quad  & \sum_{u=1}^{U} \sum_{i=1}^{N} q_i \tau_{iu} \label{objective-function} \\
  \text{s.t.} \qquad c_{iu}^{V, M},& c_{iu}^{V, S}, c_{im}^{M, M}, c_{im}^{M, S} \in \{0, 1  \},  \\
   T_{imu}^{M}, & T_{imu}^{C} \in \{0, 1  \},  \\
  \qquad \qquad  (\ref{eq1})-&(\ref{eq9}), (\ref{coupling-constraint}), (\ref{local-coupling-constraint}).  \nonumber
\end{align}
\end{subequations}

Problem (\ref{P1}) is a mixed integer nonlinear programming problem (MINLP), which is hard to solve in general. In the following, we exploit the hidden monotonicity of the problem and utilize the monotonic optimization method to solve the problem effectively.

\subsection{Monotonicity Analysis and Problem Transformation}
It can be observed that the objective function (\ref{objective-function}) monotonically increases with the increase of the caching variables $c_{iu}^{V, M}, c_{im}^{M, M}$ and the task offloading variables $ T_{imu}^{M}, T_{imu}^{C}$. However,  (\ref{objective-function}) is not a monotonic function of the power allocation variables $p_{mu}$. Nevertheless, inspired by the concept of exploiting the hidden monotonicity in the objective function using the general linear fractional programming (GLFP) \cite{zhang2013monotonic}, (\ref{objective-function}) is actually a monotonically decreasing function of the positive-valued function $\gamma_{mu}(\mathbf{p})$. More specifically, for fixed caching and task offloading variables, Problem (\ref{P1}) is equivalent to
\begin{subequations}\label{P2}
\begin{align}
	 \min_{\mathbf{y}} \quad  & \sum_{u=1}^{U} \sum_{i=1}^{N} q_i \tau_{iu} (\mathbf{y}) \label{objective-function2} \\
  \text{s.t.} \quad & \quad \mathbf{y} \in \mathcal{G},
\end{align}
\end{subequations}
where $\mathcal{G} = \{ \mathbf{y} | 0 \leq y_{mu} \leq \gamma_{mu}(\mathbf{p}), \forall m \in \mathcal{M}, \forall u \in \mathcal{U}, (\ref{eq9}) \} $.
Then, we should transform Problem (\ref{P1}) into the form of a monotonic optimization problem.

\textit{Definition (Monotonic Optimization)} \cite{tuy2000monotonic}: An optimization problem can be classified as a monotonic optimization (MO) problem if it can be written in the following canonical form,
\begin{equation}
\max _{\mathbf{x}}\{f(\mathbf{x}) \mid \mathbf{x} \in \mathcal{G} \cap \mathcal{H}\}, \mathbf{x} \in [\mathbf{a}, \mathbf{b}],
\end{equation}
where $\mathbf{x} \in \mathbb{R}_{+}^{n}$, $[\mathbf{a}, \mathbf{b}] \subset \mathbb{R}_{+}^{n}$, $f: \mathbb{R}_{+}^{n} \rightarrow \mathbb{R}$ is a monotonic increasing function of $\mathbf{x}$,
$\mathcal{G} \subset \mathbb{R}_{+}^{n}$ is a normal set with nonempty interior, and $\mathcal{H}$ is a conormal set in $[0, \mathbf{b}] $.

Based on the above monotonicity analysis of Problem (\ref{P1}), it is observed that the monotonicity of the objective function (\ref{objective-function}) with respect to the caching and task offloading variables is the opposite of that with respect to the power allocation variables. According to the definition of monotonic optimization, the objective function to be minimized should be a monotonically decreasing function with respect to the optimization variables. Thus, variable substitutions are made for the caching and task offloading variables to meet the monotonic requirements. Specifically, let $\bar{c}_{iu}^{V, M} \triangleq 1 - c_{iu}^{V, M}$, and $\bar{c}_{im}^{M, M} \triangleq 1 - c_{im}^{M, M}$ for the caching variables, and similar substitutions can be done to the task offloading variables $T_{imu}^{M}, T_{imu}^{C}$. Besides, it can be inferred that the constraints of Problem (\ref{P1}) forms the intersection of a normal set and a conormal set. Then Problem (\ref{P1}) can be equivalently transformed into a monotonic optimization problem.

\subsection{Proposed Algorithm Based on DBRB}
For the monotonic optimization problem, the most widely adopted algorithm is the polyblock outer approximation (POA) approach \cite{tuy2000monotonic}, which attempts to construct a sequence of polyblocks on the boundary of the constraints to approximate the optimal solution of continuous variables. However, Problem (\ref{P1}) cannot be solved using the POA approach since it involves the discrete caching and task offloading optimization variables. A discrete branch-reduce-and-bound (DBRB) algorithm was proposed in \cite{tuy2007discrete} to deal with the discrete variables in a monotonic optimization problem.
The DBRB algorithm draws on the idea of the standard branch-and-bound algorithm that using box\footnote{As defined in \cite{tuy2000monotonic}, a box $[\mathbf{a}, \mathbf{b}]$ is the set of  all $\mathbf{x} \in \mathbb{R}^n$ that satisfies $\mathbf{a} \leq \mathbf{x} \leq \mathbf{b}$. A box is also referred to as a  hyper-rectangle.} division based on three basic operations at each iteration: branching, reduction, and bounding, to converge to a near-optimal solution rapidly.
In this letter, we propose an improved DBRB algorithm that adapts to the specific properties of Problem (\ref{P1}).

The proposed joint caching, power control and task offloading (JCPT) algorithm based on the DBRB framework is outlined in Algorithm 1, where $B_n$, $f_{max}^{(n)}$ and $f_{min}^{(n)}$ represents the set of boxes, the upper bound and the lower bound, respectively. $r(\cdot)$ represents the reduction operation for a box.  in the $nth$ iteration.
On the basis of three basis operations at each iteration, we have made two improvements considering the properties of the problem to accelerate the solution searching process.
First, the branching and reduction operations in steps \ref{step6}-\ref{step8} can be improved considering that the caching and task offloading variables are all binary variables. Specifically, given a Box $[\mathbf{a}, \mathbf{b}] \subset \{0, 1\}^{n}$, if $b_k - a_k = 1$, we can branch on its $k$th dimension and obtain two new Box $[\mathbf{a}, \mathbf{a^{\prime}}]$ and $[\mathbf{b^{\prime}}, \mathbf{b}]$, where the value of each dimension of $\mathbf{a^{\prime}}$ is the same as that of $\mathbf{b}^{\prime}$ except that the $k$th dimension of $a_k^{\prime}$ is 0, and the $kth$ dimension $b_k^{\prime}$ is 1. Thus, the dimension of each branched box is reduced by one, which reduces the searching space. Also, the reduction operation can also be improved using the same technique.
Second, the bounding operation in step \ref{step11} can be improved considering the binary feature of the task offloading variables. In the standard DBRB framework, the bounding operation is to find a projection point of the upper right vertex of a box on the constraint curve, then the projection point serves as the lower bound of the solution, and the L-shaped area around the projection point serves as the upper bound of the solution. Nevertheless, the binary task offloading variables make the projection point not be a feasible solution. Alternatively, a heuristic algorithm can be adopted to find a tighter upper and lower bound of the solution, e.g., by randomly assigning the MEC task offloading solution and obtain the joint caching and power allocation solution using  a greedy algorithm. Thus, a feasible solution utilized to compute the upper or lower bound can be obtained.



\begin{algorithm}[htbp]
	\caption{The JCPT algorithm}
	\label{alg:1}
	\begin{algorithmic}[1]
		\STATE \textbf{Input:} $\mathbf{a}$, $\mathbf{b}$;
		\STATE \textbf{Initialization:}   $\mathcal{B}_{1} = \{ r[\mathbf{a}, \mathbf{b}]  \}$,  $n \leftarrow 1 $, $f_{min}^{(1)} \leftarrow 0$, $f_{max}^{(1)} \leftarrow + \infty $, $\gamma \leftarrow + \infty$
		\WHILE {not converged}
		\STATE $n \leftarrow n + 1;$
		\STATE  Select Box $\mathbf{B}_{s}$ that has the maximum upper bound of objective function in $\mathcal{B}_{n-1}$;
		\STATE  Branch on Box $\mathbf{B}_{s}$ to obtain Box $\mathbf{B}_{s}^{(1)}$ and $\mathbf{B}_{s}^{(2)}$;  \label{step6}
		\IF {$\mathbf{B}_{s}^{(k)}$ is feasible and contains a better solution}
		\STATE Compute $r(\mathbf{B}_{s}^{(k)})$, $k \in \{1,2\}$;  \label{step8}
		\ENDIF
	
		\IF {$r(\mathbf{B}_{s}^{(k)})$ is feasbile}
		\STATE Adopt a heuristic algorithm to compute the upper and lower bounds in this iteration; \label{step11}
		\STATE Update $f_{min}^{(n)}$ and $f_{max}^{(n)}$, $k \in \{1,2\}$;
		\ENDIF
		\IF { $f_{max}^{(n)} < \gamma$ }
		\STATE Update $\gamma \leftarrow f_{max}^{(n)}$ and record the solution $x^{*}$.
		\ENDIF
		\STATE Add the feasible box in $r(\mathbf{B}_{s}^{(1)})$ and $r(\mathbf{B}_{s}^{(2)})$ into $\mathcal{B}_{n-1}$ and remove Box $\mathbf{B}_{s}$;
		\STATE $\mathbf{B}_{n} \leftarrow \mathbf{B}_{n-1}$;
		\ENDWHILE
        \STATE \textbf{Output} $(x^{*}, \gamma)$;	
	\end{algorithmic}
\end{algorithm}

\section{Performance Evaluation} \label{Performance_Evaluation}
Simulations are carried out to validate the proposed joint HC and VC network architecture for VR delivery.
The simulations are conducted in a square area of 100 m $\times$ 100 m, where 40 SBSs and 100 HMDs are distributed in this area unless otherwise stated. There are a total of 100 VR viewpoints ranging in size from 10 Mb to 30 Mb. The maximum transmit power of SBSs is 30 dBm, and the downlink bandwidth is set to 1 GHz. For the sake of comparison, four benchmark algorithms are listed as follows. 1) Nearest Offloading  (NO): a task is offloaded to the nearest MES if the local caching or computation resources cannot meet the requirements. In other words, HC is not adopted in this algorithm; 2) Power Equally Allocation (PEA): transmit power is equally allocated among HMDs within coverage, then the caching and task offloading strategies are jointly optimized; 3) Popularity-first (PF): MESs and HMDs cache the most popular viewpoints, then the task offloading and power allocation strategies are jointly optimized; 4) Least Recently Used (LRU): a classical caching placement algorithm that substitutes the least recently requested MVs or SVs.

\begin{figure}
  \centering
  \subfigure[]{
    \label{fig:Caching-capacity-Sum-delay} 
    \includegraphics[width=4.1cm]{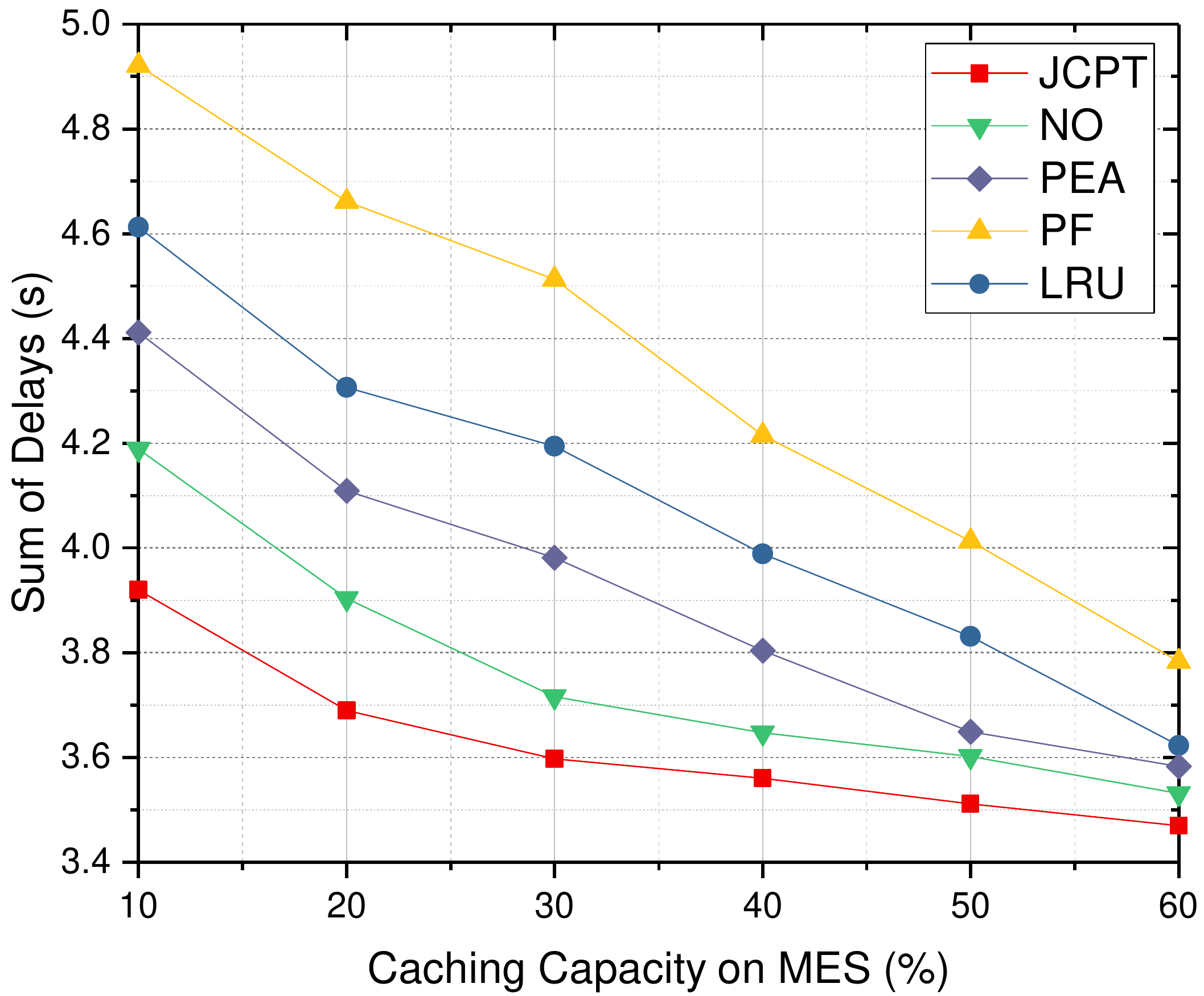}}
  \hspace{0.02in}
  \subfigure[]{
    \label{fig:SBS-number-Sum-delay} 
    \includegraphics[width=4.1cm]{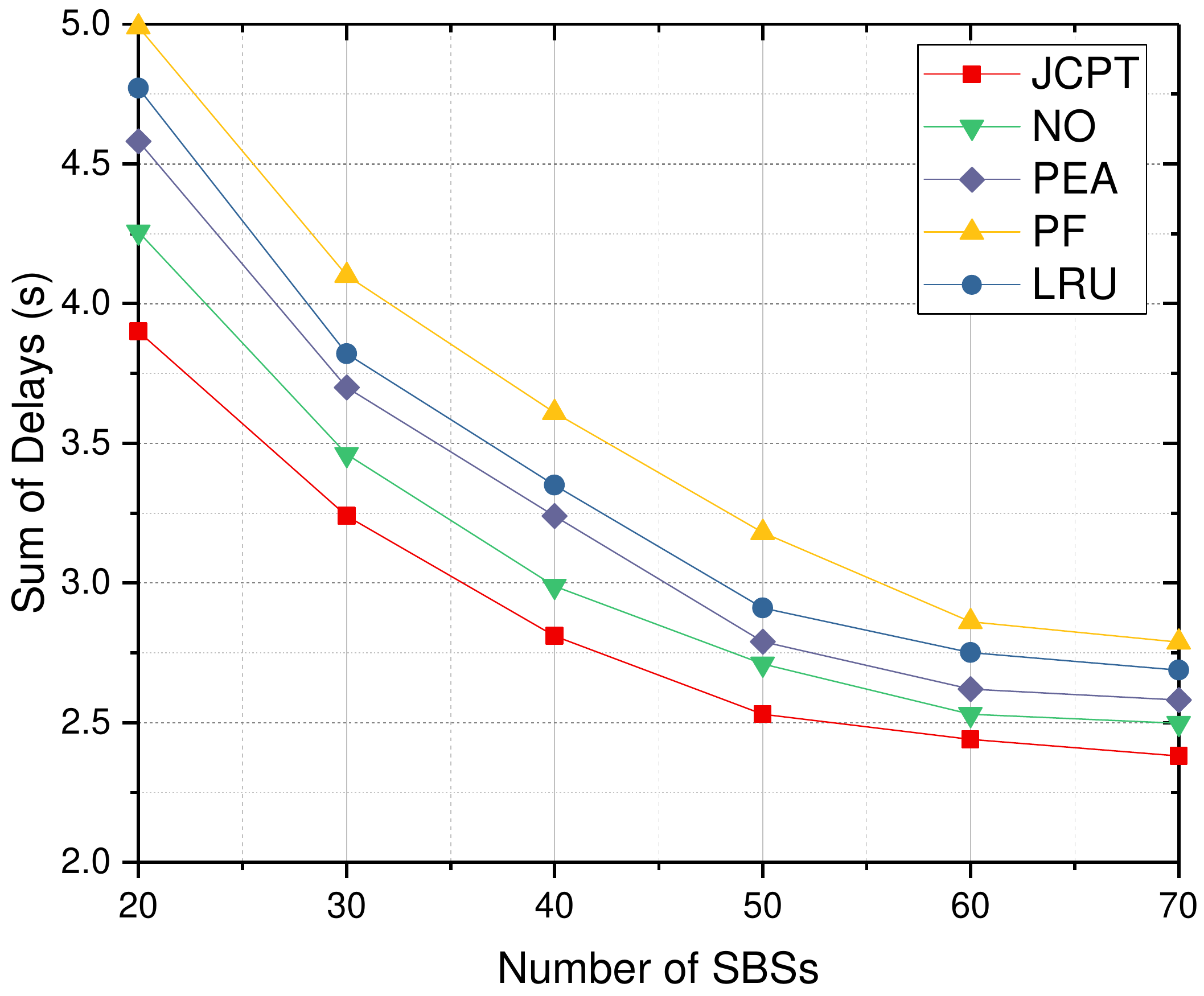}}
  \caption{Sum of delays versus (a) caching capacity on a MES, and (b) number of SBSs.}\label{fig2}
\end{figure}

Fig. \ref{fig2} shows the sum of delays with respect to the MES caching capacity and the number of SBSs using different algorithms. It is observed that the proposed JCPT algorithm achieves the lowest sum of delays in various network parameters. The NO algorithm achieves the second best delay performance, while the PF algorithm achieves the worst performance. This is because the proposed algorithm comprehensively considers the HC and VC in wireless and edge networks, especially making full use of the overlapping coverage characteristics of dense small cell networks, so that the utilization of communication, caching and computing resources for VR delivery can be maximized. In contrast, the NO algorithm is based on the classical nearby association and task offloading principles, which does not make full use of the HC among multiple MESs, and only considers VC to offload local tasks to the MESs or the cloud server, thus resulting in performance degradation. The PEA algorithm does not take into consideration the coupling relationship between wireless transmit power and heterogeneous caching and computation resources on the MESs, so the sum of delays is even higher. The PF algorithm is a greedy heuristic algorithm which does not consider the characteristics of VR videos, so it fails to obtain a better strategy in the decision of caching MVs or SVs of different viewpoints, which results in the highest the delay performance.

\begin{figure}[htbp]
  \centering
  \subfigure[]{
    \label{fig:Caching-capacity-Hit-ratio} 
    \includegraphics[width=4.1cm]{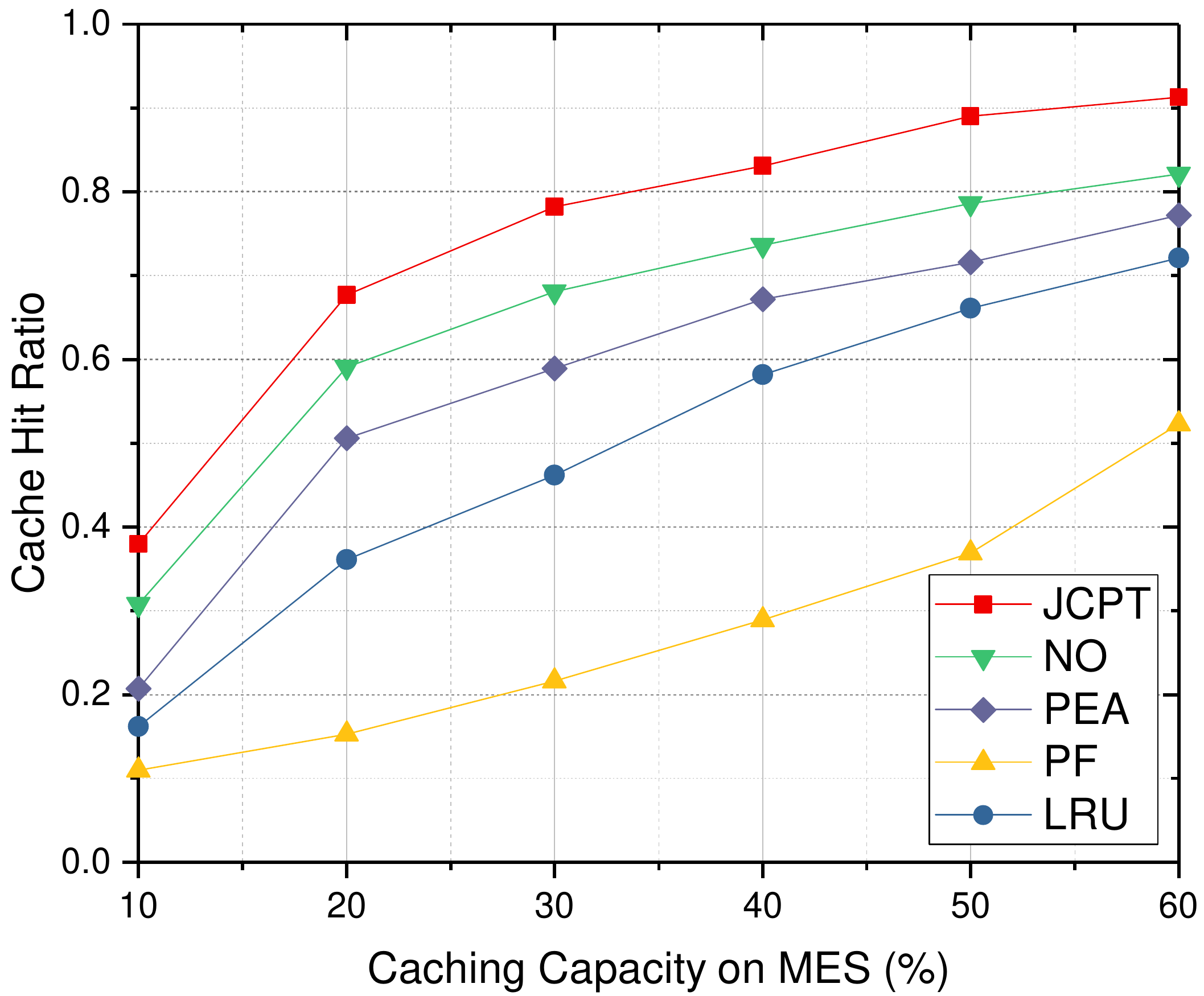}}
  \hspace{0.02in}
  \subfigure[]{
    \label{fig:Zipf-Hit-Ratio} 
    \includegraphics[width=4.1cm]{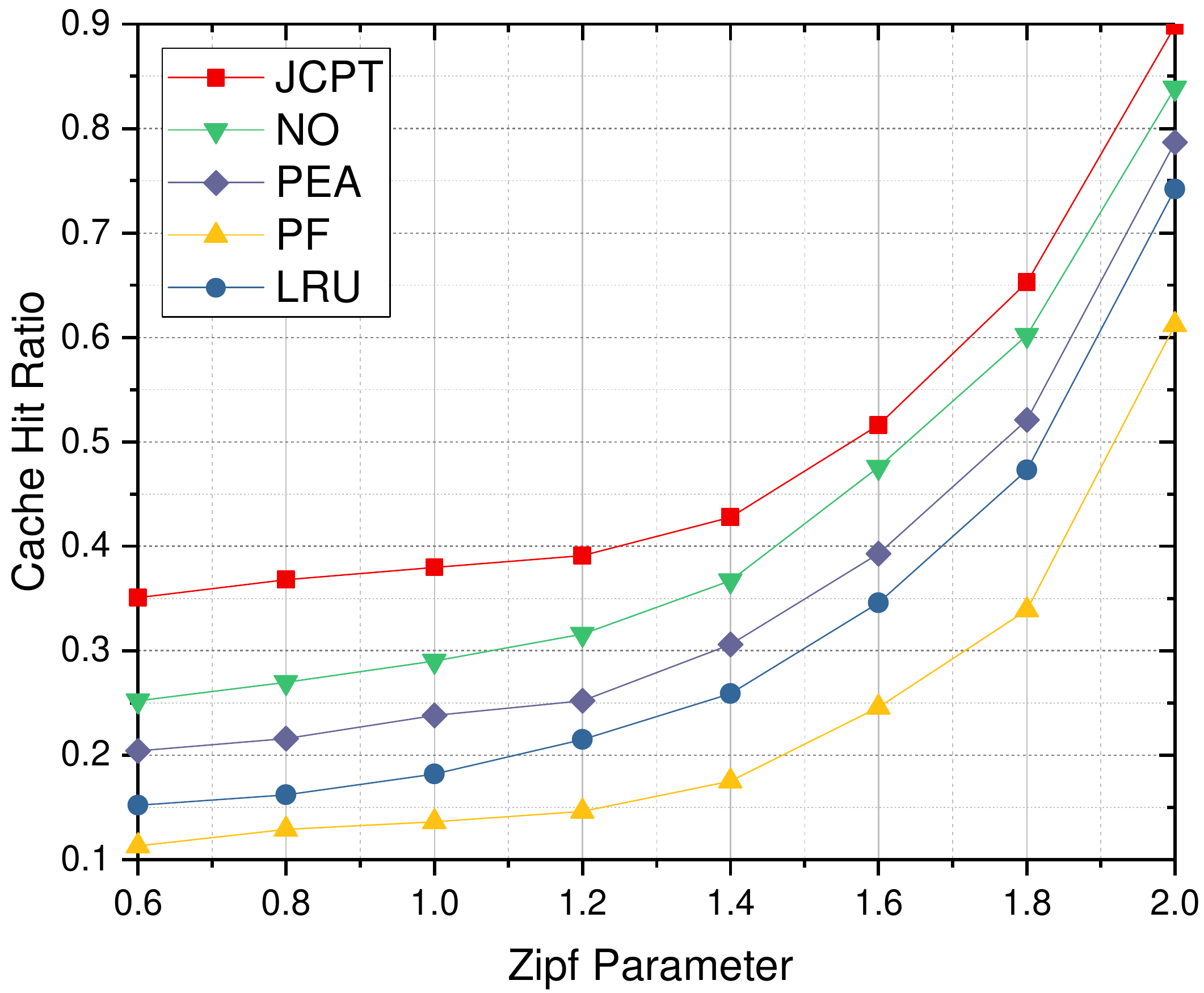}}
  \caption{Cache Hit ratio versus (a) caching capacity on a MES, and (b) the zipf parameter.} \label{fig3}
\end{figure}

Fig. \ref{fig3} shows the cache hit ratio with respect to the MES cache capacity and the zipf viewpoint popularity parameter with different algorithms. It can be seen that the proposed JCPT algorithm achieves the highest cache hit ratio. Combining with the observation in Fig. \ref{fig2}, it can be inferred that a higher cache hit ratio will lead to a lower sum of delays, which indicates the importance of an edge caching hit event to reduce the end-to-end latency of VR delivery over MEC-enabled small cell networks. Since the proposed JCPT algorithm exploits the multi-connectivity advantage of a HMD associated to multiple SBSs, the HMD can obtain the requested viewpoints directly from different MESs. Therefore, the JCPT algorithm takes advantage of this characteristic in making caching strategies, which can be viewed as a collaborative caching strategy among MESs. Meanwhile, the computation requirements of VR videos have been jointly considered with the computation resources on MESs and wireless resources of SBSs using the JCPT algorithm, thus providing a higher cache hit ratio and reducing the end-to-end latency of VR delivery.

\section{Conclusion} \label{Conclusion}

In this letter, we design a collaborative HC and VC architecture in MEC-enabled small cell networks for low-latency on-demand VR delivery. A joint horizontal and vertical task offloading model is introduced with consideration of caching and wireless communication resources to fully exploit the advantage of small cell networks for VR delivery. The hidden monotonicity of the formulated end-to-end latency minimization problem is analyzed and an improved DBRB-based algorithm is designed to efficiently solve the problem. Simulation results validate the effectiveness of the proposed JCPT algorithm in comparison with the existing benchmark algorithms and show great promise of the proposed joint HC and VC architecture for improving the quality of VR delivery over MEC-enabled small cell networks.


\begin{thebibliography}{00}
\bibitem{hosseini2017view}
M. Hosseini, ``View-aware tile-based adaptations in 360 virtual reality video streaming," in {\em 2017 IEEE Virtual Reality (VR)}, Los Angeles, CA, 2017, pp. 423--424.


\bibitem{liu2021learning}
X. Liu and Y. Deng, ``Learning-based prediction, rendering and association optimization for MEC-enabled wireless virtual reality (VR) network,'' {\em IEEE Transactions on Wireless Communications}, doi: 10.1109/TWC.2021.3073623.

\bibitem{qian2020rein}
Y. Qian, R. Wang, J. Wu, B. Tan, and H. Ren, ``Reinforcement learning-based optimal computing and caching in mobile edge network,'' {\em IEEE Journal on Selected Areas in Communications}, vol. 38, no. 10, pp. 2343--2355, 2020.

\bibitem{abbas2018mobile}
N. Abbas, Y. Zhang, A. Taherkordi, and T. Skeie, ``Mobile edge computing: A survey," {\em IEEE Internet of Things Journal}, vol. 5, no. 1, pp. 450--465, 2018.

\bibitem{sun2019communications}
Y.~Sun, Z.~Chen, M.~Tao, and H.~Liu, ``Communications, caching, and computing for mobile virtual reality: Modeling and tradeoff,'' \emph{IEEE Transactions on Communications}, vol.~67, no.~11, pp. 7573--7586, 2019.

\bibitem{dang2019joint}
T.~Dang and M.~Peng, ``Joint radio communication, caching, and computing design for mobile virtual reality delivery in fog radio access networks,''  \emph{IEEE Journal on Selected Areas in Communications}, vol.~37, no.~7, pp. 1594--1607, 2019.

\bibitem{wu2019delay}
H. Wu and H. Lu, ``Delay and power tradeoff with consideration of caching capabilities in dense wireless networks,'' {\em IEEE Transactions on Wireless Communications}, vol. 18, no. 10, pp. 5011--5025, 2019.


\bibitem{gu2021association}
Z. Gu, H. Lu, M. Zhang, H. Sun, and C. W. Chen, ``Association and caching in relay-assisted mmWave networks: From a stochastic geometry perspective,'' {\em IEEE Transactions on Wireless Communications}, doi: 10.1109/TWC.2021.3091815.


\bibitem{wang2018enabling}
K. Wang, H. Yin, W. Quan, and G. Min, ``Enabling collaborative edge computing for software defined vehicular networks," {\em IEEE Network}, vol. 32, no. 5, pp. 112--117, 2018.

\bibitem{saputra2021anovel}
Y. M. Saputra, D. T. Hoang, D. N. Nguyen, and E. Dutkiewicz, ``A novel mobile edge network architecture with joint caching-delivering and horizontal cooperation,'' {\em IEEE Transactions on Mobile Computing}, vol. 20, no. 1, pp. 19--31, 2021.

\bibitem{dai2020a}
P. Dai, K. Hu, X. Wu, H. Xing, F. Teng, and Z. Yu, ``A probabilistic approach for cooperative computation offloading in MEC-assisted vehicular networks,'' {\em IEEE Transactions on Intelligent Transportation Systems}, pp. 1--13, 2020.

\bibitem{3GPP}
3GPP, ``Evolved universal terrestrial radio access (E-UTRA) and NR; Multi-connectivity,'' (Release 16), 3GPP TS 37.340, V16.6.0, 2021.

\bibitem{wolf2019how}
A. Wolf, P. Schulz, M. Dorpinghaus, J. C. S. S. Filho, and G. Fettweis, ``How reliable and capable is multi-connectivity?,'' {\em IEEE Transactions on Communications}, vol. 67, no. 2, pp. 1506--1520, 2019.

\bibitem{mao2017a}
Y. Mao, C. You, J. Zhang, K. Huang, and K. B. Letaief, ``A survey on mobile edge computing: The communication perspective," {\em IEEE Communications Surveys $\&$ Tutorials}, vol. 19, no. 4, pp. 2322--2358, 2017.

\bibitem{zhang2013monotonic}
Y. J. A. Zhang, L. Qian, and J. Huang, ``Monotonic optimization in communication and networking systems,'' {\em  Foundations and Trends in Networking}, vol. 7, no. 1, pp. 1--75, 2013.

\bibitem{tuy2000monotonic}
H. Tuy, ``Monotonic optimization: Problems and solution approaches,'' {\em SIAM Journal on Optimization}, vol. 11, no. 2, pp. 464--494, 2000.

\bibitem{tuy2007discrete}
H. Tuy, M. Minoux, and N. T. Hoai-Phuong, ``Discrete monotonic optimization with application to a discrete location problem,'' {\em SIAM Journal on Optimization}, vol. 17, no. 1, pp. 78--97, 2007.


	

\end{thebibliography}
\end{document}